\documentclass[11pt]{article} 
\usepackage{hyperref} 
\pdfoutput=1 
 
\begin{document} 
 
\title{Hydrodynamic causes and effects of air bubbles rising in very viscous media \\ Fluid Dynamics Videos} 
 
\author{Sharad Chand Ravinuthala \\ 
\\\vspace{6pt} Department of Mechanical Engineering
 \\ West Virginia University, Morgantown, WV 26505, USA} 
 
\maketitle 
 
%% The abstract (in this file, and that submitted as text to arXiv) should 
%% include the exact phrase 
%% "fluid dynamics video" or "fluid dynamics videos" 
 
\begin{abstract} 
The current fluid dynamics video, showcases the hydrodynamic behaviour of bubbles in a viscous media as
compared with a medium of lesser viscosity. The level of mixing that the two-phase gas in liquid
flow causes is the parameter of interest 
\end{abstract} 
 
% main text 
 
\section{Introduction} 
 
The {\em hyperref} package is used to make links to the videos. 
%% The format is: \href{URL of video}{name that will appear in the text} 
 
Two sample videos are 
\href{http://ecommons.library.cornell.edu/bitstream/1813/8237/2/LIFTED_H2_EMS
T_FUEL.mpg}{Video 
1} and 
\href{http://ecommons.library.cornell.edu/bitstream/1813/8237/4/LIFTED_H2_IEM
_FUEL.mpg}{Video 
2}. 
 
Detailed understanding of two-phase gas liquid flows is imperative for developing 
efficient multi-phase reactors through precise control of mixing and reaction
kinetics. The bubble column is a good apparatus for elementary studies of such 
flows. In the current study experiments are conducted to assess the effect of
liquid viscosity on flow dynamics inside a bubble column. Corn oil and water 
are used as the continuous media, and air was the dispersed media. The objective
of this effort is to use the results for a qualitative validation of the numerical
simulations.

\end{document}